\documentclass[final]{elsart}

\usepackage{amssymb}
\usepackage{graphicx}

\newcommand{\thsp}{\thinspace}

\newcommand{\alp}{\ensuremath{\alpha}}
\newcommand{\mnuc}[2]{\ensuremath{\mathrm {^{#2}#1}}}
\newcommand{\eg}{e.g.}

\newcommand{\edot}{\ensuremath {\dot \epsilon}}

\newcommand{\sigv}{{\ensuremath{\langle \sigma v \rangle}}}
\newcommand{\ye}{\ensuremath{Y_{e}}}
\newcommand{\calN}{\ensuremath {\mathcal N}}
\newcommand{\calF} {\ensuremath {\mathcal {F}}}
\newcommand{\calZ} {\ensuremath {\mathcal {Z}}}
\newcommand{\calR} {\ensuremath {\mathcal {R}}}

\newcommand{\calG}{\ensuremath {\mathcal {G}}}

\newcommand{\yf}{\ensuremath {\vec {Y}^{\calF}}}
\newcommand{\yr}{\ensuremath {\vec {Y}^{\calR}}}

\newcommand{\yg}{\ensuremath {\vec {Y}^{\calG}}}
\newcommand{\ygdot}{\ensuremath {\dot {{\vec Y}^{\calG}}}}
\newcommand{\mev}{\ensuremath{\mathrm{\thsp MeV}}}

\newcommand{\cc}{\ensuremath {\mathrm{cm^{3}}}}
\newcommand{\gcc}{\ensuremath{\mathrm{\thsp g \thsp cm^{-3}}}}

\newcommand{\gk}{\ensuremath{\mathrm{\thsp GK}}}
 
\newcommand{\pergm}{\ensuremath{\mathrm{\thsp g^{-1}}}}
\newcommand{\persec}{\ensuremath{\mathrm{\thsp s^{-1}}}}     

\begin{document}

\begin{frontmatter}



\title{Thermonuclear Kinetics in Astrophysics}

\author[AUT1,AUT2]{W. Raphael Hix}
\author[AUT3]{Bradley S. Meyer}

\address[AUT1]{Physics Division, Oak Ridge National Laboratory, Oak Ridge, TN 37831}
\address[AUT2]{Department of Physics \& Astronomy, University of
Tennessee, Knoxville, TN 37996}
\address[AUT3]{Department of Physics \& Astronomy, Clemson University, Clemson, SC 29634}  

\begin{abstract}
% Text of abstract
Over the billions of years since the Big Bang, the lives, deaths and afterlives of stars have enriched the Universe in the heavy elements that make up so much of ourselves and our world.  This review summarizes the methods used to evolve these nuclear abundances within astrophysical simulations.  These methods fall into 2 categories;  evolution via rate equations and via equilibria.  Because the rate equations in nucleosynthetic applications involve a wide range of timescales, implicit methods have proven mandatory, leading to the need to solve matrix equations.  Efforts to improve the performance of such rate equation methods are focused on efficient solution of these matrix equations, in particular by making best use of the sparseness of these matrices, and finding methods that require less frequent matrix solutions.  Recent work to produce hybrid schemes which use local equilibria to reduce the computational cost of the rate equations is
also discussed.  Such schemes offer significant improvements in the speed of
reaction networks and are accurate under circumstances where calculations which assume complete equilibrium fail. 
\end{abstract}

\end{frontmatter}

% main text

\section{Introduction} \label{sect:intro}

Research by Helmholtz, Kelvin and others throughout the second half of the 19th century made it clear that neither gravity nor any other then known energy 
source could account for the geologically determined age of the Sun and solar system.  Enlightened by Rutherford's 1911 discovery of the atomic nucleus, Eddington and others suggested that nuclear transmutations might be the remedy to this quandary.  With the burgeoning knowledge of the properties of nuclei and nuclear reactions in the 1930s, 1940s and 1950s came a growing understanding of the role that individual nuclear reaction played in the synthesis of the elements.  In 1957, Burbidge, Burbidge, Fowler \& Hoyle \cite{BBFH57} and Cameron \cite{Came57} wove these threads into a cohesive theory of nucleosynthesis, and demonstrated how the solar isotopic abundances bore the fingerprints of their astrophysical origins.  Today, investigations refine our answers to these same two questions; how are the elements we see on Earth and throughout the universe formed, and how do these nuclear transmutations, and the energy they release, affect their astrophysical hosts?

In this article, we will review the numerical methods used to study these questions.  Two basic numerical methods are at the heart of the study of thermonuclear kinetics in astrophysics, the tracking of nuclear  transmutations via rate equations and via equilibria.  We will also briefly discuss work which seeks to meld these methods together in order to overcome the limitations of each.  Additionally we will discuss the issues encountered when coupling the nuclear evolution to hydrodynamic simulations.

\section{Thermonuclear Reaction Networks} \label{sect:network}

Composed of a system of first order differential equations, the 
nuclear reaction network has sink and source terms representing each of the 
many nuclear reactions involved.  Prior to discussing the numerical 
difficulties posed by the nuclear reaction network, it is necessary to understand 
the sets of equations we are attempting to solve.  To this end, we present 
a brief overview of the thermonuclear reaction rates of interest and how 
these rates are assembled into the differential equations that must ultimately be solved.  For more detailed information, we refer the reader to several textbooks covering this subjects \cite{Clay83,RoRo88,Arne96}.  We will end this section by briefly discussing the coupling of nucleosynthesis with hydrodynamics.

\subsection{Thermonuclear Reaction Rates} \label{sect:reacrate}

There are a large number of types of nuclear reactions which are of 
astrophysical interest.  In addition to the emission or absorption of  nuclei
and nucleons, nuclear reactions can involve the emission or  absorption of
photons ($\gamma$-rays) and leptons (electrons, neutrinos,  and their
anti-particles).  As a result, nuclear reactions involve three of the  four
fundamental forces, the nuclear strong, electromagnetic and nuclear weak 
forces.  Reactions involving leptons (termed weak interactions) proceed  much
more slowly than those involving only nucleons and photons;  however,  these
reactions are important because only weak interactions can change the global
ratio of protons to neutrons.

The most basic piece of information about any nuclear reaction is the
nuclear cross section.  The cross section for a reaction between target 
$j$ and projectile $k$ is defined by 
\begin{equation}
    \sigma = {\rm{number\ of\ reactions\ target^{-1} sec^{-1}} \over
    {flux\ of\ incoming\ projectiles}} = {{r/n_j} \over {n_k v}}. 
    \label{eq:sigma}
\end{equation}
The second equality holds when the relative velocity between targets of 
number density $n_j$ and projectiles of number density $n_k$ is constant 
and has the value $v$.  Then $r$, the number of reactions per \cc\ and sec, 
can be expressed as $r=\sigma v n_j n_k$.  More generally, the targets and 
projectiles have distributions of velocities, in which case $r$ is given by
\begin{equation}
    r_{j,k}=\int \sigma (\vert \vec v_j -\vec v_k\vert) \vert \vec v_j 
    -\vec v_k\vert d^3 n_j d^3 n_k.
    \label{eq:rate}
\end{equation}
The evaluation of this integral depends on the types of particles and
distributions which are involved. For nuclei $j$ and $k$ in an astrophysical 
plasma, Maxwell-Boltzmann statistics generally apply; thus,
\begin{equation}
    d^3n=n ({{m} \over {2\pi k_B T}})^{3/2} \exp (-{{mv^2} \over {2 k_B T}}) d^3v,
\end{equation}

allowing $n_j$ and $n_k$ to be moved outside of the integral.  
Eq.~\ref{eq:rate} can then be written as $r_{j,k}=\sigv_{j,k} n_j n_k$, 
where \sigv\ is the velocity integrated cross section.  Equivalently, one 
can express the reaction rate in terms of a mean lifetime of particle $j$
against destruction by particle $k$, 
\begin{equation}
    \tau_{k}(j)= {1 \over {\sigv_{j,k} n_{k}}}
    \label{eq:tau}
\end{equation}
For thermonuclear reactions, these integrated cross sections have the form 
\cite{Clay83,FoCZ67}

\begin{equation}
    \langle j,k \rangle \equiv \sigv_{j,k}= ({8 \over {\mu \pi}})^{1/2} 
    (k_B T)^{-3/2} \int_0 ^\infty E \sigma (E) {\rm exp}(-E/k_B T) dE, 
    \label{eq:sigv}
\end{equation}
where $\mu$ denotes the reduced mass of the target-projectile system, $E$ 
the center of mass energy, $T$ the temperature and $k_{B}$ is Boltzmann's 
constant.

Experimental measurements and theoretical predictions for these reaction rates
provide the data input necessary to study astrophysical thermonuclear kinetics.  While detailed discussion of individual rates is beyond the scope of this article, the interested reader is directed to the following reviews, in this volume and elsewhere.  Experimental nuclear rates and the methodology for measuring them have been reviewed in detail by \cite{RoRo88,KaTW98}.  A number of articles in this volume review the charged particle reaction rates which are important in the Sun \cite{HaPR04}, in later stages of stellar evolution \cite{BuBa04}, and for explosive burning on compact objects \cite{AnSB04,HeJI04,ReSc04}.  Experimental neutron capture cross sections are also summarized \cite{KaMe04,GeKS04}.  Theoretical modeling of these rates \cite{RaDe04,MaNW04} is vitally important to provide the many rates for which experimental information is incomplete or non-existant.  Beyond measuring or calculating these rates, progress in nuclear astrophysics also requires the compilation and dissemination (in a usable form) of this hard-earned nuclear data to the broadest astrophysical community \cite{CaFo88,GRSD04,Smit03}. 

When particle $k$ in Eq.~\ref{eq:rate} is a photon, the distribution 
$d^3n_k$ is given by the Plank distribution. 
%\begin{equation}
%    d^3n_{\gamma} =  {{8 \pi} \over {c^3 h^3}} {{E_{\gamma}^2} \over
%    {\exp \left(E_{\gamma} /k_B T\right)-1}} dE_{\gamma} \ .
%\end{equation}
Furthermore, the relative velocity is always $c$ and thus the integral 
is separable, simplifying to 
\begin{equation}
    r_j  = {{\int d^3n_j}\over {\pi ^2 (c \hbar )^3}} \int_0 ^\infty {{c 
    \sigma(E_\gamma ) E_{\gamma}^2} \over {{\rm exp}(E_\gamma /k_B T) -1}} dE_
    \gamma \equiv \lambda_{j,\gamma} (T) n_j. 
    \label{eq:phrate}
\end{equation}
In practice it is not usually necessary to directly evaluate the 
photodisintegration cross sections (see, however, \cite{MoZU04} for exceptions), because they can be expressed by detailed balance in terms of the capture cross sections for the inverse 
reaction, $l+m\rightarrow j+\gamma$ \cite{FoCZ67} .
\begin{equation}
    \lambda_{j,\gamma} (T)= ({{G_l G_m} \over G_j}) ({{A_l A_m} \over A_j})
    ^{3/2} ({{m_u k_B T} \over {2\pi \hbar^2}})^{3/2} \langle l,m \rangle 
    \exp (-Q_{lm}/k_B T).
    \label{eq:detbal}
\end{equation}
This expression depends on the partition functions, $G_{k}=\sum_i (2J_i+1) 
\exp (-E_i/k_B T)$ (which account for the populations of the excited 
states of the nucleus), the mass numbers, $A$, the temperature 
$T$, the inverse reaction rate $\langle l,m \rangle$, and the reaction 
$Q$-value (the energy released by the reaction), $Q_{lm}=(m_l+m_m-m_j) 
c^2$.  Since photodisintegrations are endoergic, their rates are 
vanishingly small until sufficient photons exist in the high energy tail of 
the Planck distribution with energies $> Q_{{lm}}$.  As a rule of thumb this 
requires $T$$\approx$$Q_{lm}/30 k_{B}$.

In practice, these experimental and theoretical reaction rates are determined 
for bare nuclei, while in astrophysical plasmas, these reactions occur 
among a background of other nuclei and electrons.  As a result of this 
background, the reacting nuclei experience a Coulomb repulsion modified from 
that of bare nuclei. For high densities and/or low temperatures, the effects 
of this screening of reactions becomes very important. Under most 
conditions (with non-vanishing temperatures) the generalized reaction rate 
integral can be separated into the traditional expression without screening 
[Eq.~\ref{eq:sigv}] and a screening factor, 
\begin{equation}
    \langle j,k \rangle^*=f_{scr}(Z_j,Z_k,\rho,T,n_i) \langle j,k \rangle. 
    \label{eq:screen}
\end{equation}
This screening factor is dependent on the charge of the involved particles,
the density, temperature, and the composition of the plasma. For more details 
on the form of $f_{scr}$, see, \eg, \cite{Salp54,DeGC73,Ichi93,BrSa97}.  
At high densities and low temperatures screening factors can enhance 
reactions by many orders of magnitude and lead to {\em pycnonuclear ignition}.
In the extreme case of very low temperatures, where reactions are only
possible via ground state oscillations of the nuclei in a Coulomb lattice,
Eq.~\ref{eq:screen} breaks down, because it was derived under the assumption 
of a Boltzmann distribution \cite{SaVa69,IcKi99}.

In stellar plasmas, target nuclei also do not exist solely in their ground
states.  This complicates the rate expression in Eq.~\ref{eq:sigv},
which now must take into account the cross sections for capture out of the
different excited states and properly weight them according to their
probability of occurrence in the ensemble of target nuclei.  Because the
timescales for transitions between excited states of a nucleus are typically
much shorter than other reaction timescales, it is usually valid to assume
that the nuclei are internally equilibrated and a given excited state is
populated in the ensemble by the usual Boltzmann factor $\propto e^{-E/k_BT}$,
where $E$ now is the excitation energy of that state.  From this, one may
derive a factor, called the {\em stellar enhancement factor} (SEF), to
correct the ground-state reaction rate for the population of excited
states (see, e.g., \cite{RaTh00,AARD99}).

Interesting complications to this prescription arise when the internal
equilibration timescale for a nucleus exceeds the other reaction
timescales in the problem.  This usually occurs when a large spin
difference between the ground state of an isotope and its first excited state
prevents them from communicating directly via internal transitions.
The isotope $^{26}$Al is the prototypical example, with a ground and isomeric state of spin and parity $5^+$ and $0^+$, respectively.  When these two states equilibrate, it is not through a direct transition between the states--that simply takes too long.  Rather the equilibration occurs
through transitions to higher-lying levels that communicate effectively
with both levels.  This problem has been studied in detail (e.g.,
\cite{WaFo80}), and the most straightforward way of dealing with this is
to assume that the higher-lying levels are in a steady state since their
destruction timescales are so rapid.  When this is done, the population of
the target nucleus break up into two ensembles, one tied to the ground
state and one tied to the isomer \cite{GuMe01}.  These may then be treated
as separate species in the network, each with its own rates to and from
other isotopes and with effective rates for transitions between the two
ensembles.  Fortunately, the number of nuclei requiring such treatment is
small and does not add much computational burden to running the nuclear
reaction network.

A procedure similar to that used to derive Eq.~\ref{eq:phrate} applies to captures of electrons by nuclei.  Because the electron is 1836 times less massive than a 
nucleon, the velocity of the nucleus $j$ in the center of mass system is 
negligible in comparison to the electron velocity ($\vert \vec v_j- \vec 
v_e \vert \approx \vert \vec v_e \vert$).  In the neutral, completely 
ionized plasmas typical of the astrophysical sites of nucleosynthesis, the 
electron number density, $n_{e}$, is equal to the total density of protons 
in nuclei, $\sum_i Z_i n_i$.  However in many of these astrophysical 
settings the electrons are at least partially degenerate, therefore the 
electron distribution cannot be assumed to be Maxwellian.  Instead the 
capture cross section may be integrated over a Boltzmann, partially 
degenerate, or degenerate Fermi distribution of electrons, depending on the 
astrophysical conditions.  The resulting electron capture rates are 
functions of $T$ and $n_e$, $r_j=\lambda_{j,e} (T,n_e) n_j.$  Similar equations apply for the capture of positrons which are in thermal equilibrium with photons, electrons, and nuclei.  Electron and positron capture calculations have been performed for a large variety of nuclei with mass numbers between A=20 and A=100 (see \cite{LiMF04} for more information). 
For normal decays, like beta or alpha decays, with a characteristic  half-life
$\tau_{1/2}$,  Eq.~\ref{eq:phrate} also applies, with the
decay constant  $\lambda_j=\ln 2/\tau_{1/2}$.  In addition to innumerable
experimental half-life determinations, beta-decay half-lives for a wide range of unstable nuclei have  been predicted (see \cite{Borz04,TBFK04}).

Even though the size of the neutrino scattering cross section on nuclei and electrons is very small, at high densities ($\rho \sim 10^{13} \gcc$), enough scattering events occur to thermalize the neutrino distribution.  Under such conditions the inverse process to electron capture (neutrino capture) can occur in significant numbers and the neutrino capture rate can be expressed in a form similar to Eqs.~\ref{eq:phrate} by integrating over the thermal 
neutrino distribution (e.g. \cite{FuMe95}).  Inelastic neutrino scattering 
on nuclei can also be expressed in this form.  The latter can cause 
particle emission, similar to photodisintegration.  In this volume, see \cite{BuRe04,FuRa04,Voge04} for more complete discussion of the interactions of neutrinos with matter.  The calculation of these rates 
can be further complicated by the neutrinos being out of thermal 
equilibrium with the local environment.  When thermal equilibrium among 
neutrinos was established at a different location, then the neutrino 
distribution might be characterized by a chemical potential and a 
temperature different from the local values.  Otherwise, the neutrino 
distribution must be evolved in detail (see, \eg, \cite{MeMe99}). 

\subsection{Thermonuclear Rate Equations} \label{sect:yderiv}

The large number of reaction types discussed in \S \ref{sect:reacrate} can 
be divided into 3 functional categories based on the number of reactants 
which are nuclei.  The reactions involving a single nucleus, which include 
decays, electron and positron captures, photodisintegrations, and neutrino 
induced reactions, depend on the number density of only the target species.  
For reaction involving two nuclei, the reaction rate depends on the number 
densities of both target and projectile nuclei.  There are also a few 
important three-particle process (like the triple-\alp\ process, see \S \ref{sect:hyb}) which are commonly successive captures with an 
intermediate unstable target (see, \eg, \cite{NoTM85,GoWT95}).  Using an 
equilibrium abundance for the unstable intermediate, the contributions of 
these reactions are commonly written in the form of a three-particle 
processes, depending on a trio of number densities.  Grouping reactions by 
these 3 functional categories, the time derivatives of the number densities 
of each nuclear species in an astrophysical plasma can be written in terms 
of the reaction rates, $r$, as
\begin{equation}
    \left. {{\partial n_i} \over {\partial t}} \right|_{\rho =const}= \sum_j 
    \calN^i _j r_j + \sum_{j,k} \calN^i _{j,k} r_{j,k} + \sum_{j,k,l} 
    \calN^i _{j,k,l} r_{j,k,l},
    \label{eq:ndot}
\end{equation}
where the three sums are over reactions which produce or destroy a nucleus 
of species $i$ with 1, 2 \& 3 reactant nuclei, respectively.  The \calN\ s
provide for proper accounting of numbers of nuclei and are given by: 
$\calN^i_j = N_i$, $\calN^i_{j,k} = N_i / \prod_{m=1}^{n_{j,k}} | N_{m} |!  
$, and $\calN^i_{j,k,l} = N_i / \prod_{m=1}^{n_{j,k,l}} |N_{m}|!$.  The $N_i's$ 
can be positive or negative numbers that specify how many particles of 
species $i$ are created or destroyed in a reaction, while the denominators, 
including factorials, run over the $n_{j,k}$ or $n_{j,k,l}$ different species destroyed in the reaction and avoid double counting of the number of reactions when identical particles react with each other (for example in the \mnuc{C}{12} + \mnuc{C}{12} or the triple-\alp\ reactions; for details see \cite{FoCZ67}).

In addition to nuclear reactions, expansion or contraction of the plasma 
can also produce changes in the number densities $n_i$.  To separate the 
nuclear changes in composition from these hydrodynamic effects, the nuclear abundance $Y_i =n_i/\rho N_A$, where $N_{A}$ is Avogadro's number, is commonly used.  For a nucleus with atomic weight $A_i$, $A_iY_i$ 
represents the mass fraction of this nucleus, therefore $\sum A_iY_i=1$.  
Likewise, the equation of charge conservation becomes $\sum Z_i Y_i = Y_e$, 
where $\ye (= n_e /\rho N_A)$ is the electron abundance (also termed the electron fraction).  By recasting Eq.~\ref{eq:ndot} in terms of nuclear abundances $Y_i$, a set of ordinary differential equations 
for the evolution of $\dot Y_i$ results which depends only on nuclear 
reactions.  In terms of the reaction cross sections introduced in \S 
\ref{sect:reacrate}, this reaction network is described by the following 
set of differential equations
%\begin{eqnarray}
%    \dot Y_i & = & \sum_j \calN^i _j \lambda_j Y_j + \sum_{j,k} \calN^i _{j,k} 
%    \rho N_A \langle j,k \rangle Y_j Y_k  \nonumber \\ 
%    & + & \sum_{j,k,l} \calN^i _{j,k,l} \rho^2 N_A^2 \langle j,k,l \rangle 
%	Y_j Y_k Y_l. 
%    \label{eq:ydot}
%\end{eqnarray}
\begin{equation}
    \dot Y_i = \sum_j \calN^i _j \lambda_j Y_j + \sum_{j,k} \calN^i _{j,k} 
\rho N_A \langle j,k \rangle Y_j Y_k  + \sum_{j,k,l} \calN^i _{j,k,l} \rho^2 N_A^2 \langle j,k,l \rangle Y_j Y_k Y_l. \label{eq:ydot}
\end{equation}

\subsection{Coupling Nuclear Reaction Networks to Hydrodynamics}\label{sect:hydro}

As we touched on in the previous section, nuclear processes are tightly linked
to the hydrodynamic behavior of the bulk medium.  Thermonuclear  processes
release (or absorb) energy, which alters the pressure and causes hydrodynamic
motions.  These motions may disperse the thermonuclear ash and bring a
continued supply of fuel to support the thermonuclear flame.  The compositional changes caused by thermonuclear reactions can also
change the equation of state and opacity, further impacting the hydrodynamic
behavior.  For purposes of this review of thermonuclear kinetics methods, which
generally assume that thermonuclear and hydrodynamic changes in local
composition can be successfully decoupled (or treated in an operator split 
fashion), we include only a brief description of how this decoupling is best
achieved.   M\"uller \cite{Muel98} provides an authoritative overview and discusses
the difficulties (and open issues) involved when including nucleosynthesis within
hydrodynamic simulations.  

The coupling between thermonuclear processes and hydrodynamic changes can 
be divided into two categories by considering the spatial extent of the 
coupling.  Nucleosynthetic changes in composition and the resultant energy 
release produce \emph{local} changes in hydrodynamic quantities like 
pressure and temperature.  The strongest of these local couplings is the 
release (or absorption) of energy and the resultant change in temperature.  
Changes in temperature are particularly important because of the 
exponential nature of the temperature dependence of thermonuclear reaction 
rates.  Since the nuclear energy release is uniquely determined by the 
abundance changes, the rate of thermonuclear energy release, $\dot 
\epsilon$, is given by
\begin{equation}
    \dot \epsilon_{nuc} = - \sum_i  N_A M_i c^2 \dot Y_i \thinspace (\mev \pergm \persec).
    \label{eq:edot}
\end{equation}
where $M_i c^2$ is the rest mass energy of species $i$ in \mev.  Since all 
reactions conserve nucleon number, the atomic mass excess $M_{ex,i}=M_i - 
A_i m_u$ ($m_u$ is the atomic mass unit) can be used in place of the mass 
$M_i$ in Eq.~\ref{eq:edot} (see \cite{AuWa95} for a recent compilation of 
mass excesses).  The use of atomic mass units has the added benefit that 
electron conservation is correctly accounted for in the case of $\beta^{-}$ 
decays and $e^{-}$ captures, though reactions involving positrons require 
careful treatment.  In general, the nuclear energy release is deposited 
locally, so the rate of thermonuclear energy release is equal to the 
nuclear portion of the hydrodynamic heating rate.  However, there are 
instances where nuclear products do not deposit their energy locally.  
Escaping neutrinos can carry away a portion of the thermonuclear energy 
release.  In the rarefied environment of supernova ejecta at late times, 
positrons and gamma rays released by $\beta$ decays are not completely 
trapped. In most such cases, the escaping particles stream freely 
from the reaction site, allowing adoption of a simple loss term analogous to 
Eq.~\ref{eq:edot} with $M_i c^2$ replaced by an averaged energy loss term.  
For this reason, weak reaction rate tabulations provide averaged neutrino losses.  From these we can construct
\begin{equation}
    \dot \epsilon_{\nu \ loss} =  \sum_i \langle E_{\nu} \rangle \dot 
    Y_{i,weak} ,
    \label{eq:enudot}
\end{equation}
where we consider only those contributions to $\dot Y$ due to neutrino 
producing reactions.  In some cases, like supernovae, subsequent interactions between the escaping leptons or gamma rays and matter require more complete transport to be considered.  Other important quantities which 
are impacted by nucleosynthesis, like \ye, can be obtained by appropriate 
sums over the abundances and also need not be evolved separately.

Implicit solution methods require the calculation of $\dot Y (t+\Delta t)$, 
where $\Delta t$ is the nuclear timestep, which in turn requires knowledge  of
$T(t+\Delta t)$.  One could write a differential equation for the energy 
release analogous to Eq.~\ref{eq:ydot}, with the \calN s replaced by the 
reaction $Q$-values, and thereby evolve the energy release (and calculate 
temperature changes) as an additional equation within the network solution.  
M\"uller \cite{Muel86} has shown that such a scheme can help avoid 
instabilities in the case of a physically isolated zone entering or leaving 
nuclear statistical equilibrium.  In general, however, use of this additional
equation is made unnecessary by the relative slowness with which the
temperature changes.  The timescale on which the temperature changes is given by
\begin{equation}
    \tau_{T}= T / {\dot T} \approx C_{V}T/ \dot \epsilon_{nuc}
    \label{eq:taut}
\end{equation}
and is often called the \emph{ignition timescale}.  The timescale on which an 
individual abundance changes is its \emph{burning time}, 
\begin{equation}
    \tau (\mnuc{Z}{A}) = Y(\mnuc{Z}{A}) /{\dot Y(\mnuc{Z}{A})} \sim \min_{k} 
    {\tau_{k}(\mnuc{Z}{A})} 
    \label{eq:tauy}
\end{equation}
where $\tau_{k}(\mnuc{Z}{A})$ is defined in Eq.~\ref{eq:tau}.  In general 
$\tau_{T}$ differs from $\tau (\mnuc{Z}{A})$ of the principle fuel by the 
ratio of thermal energy content to the energy released by the reaction.  
For degenerate matter this ratio can approach zero, allowing for explosive 
burning.  In contrast, accurate prediction of less abundant, but still 
important, species requires that the reaction network timestep $\Delta t$ be chosen to 
be the burning timescale of a less abundant species, typically with an abundance 
of $10^{-6}$ or smaller \cite{ArTr69}.  Since the dominant fuel is most often 
one of the more abundant constituents and the burning timescales are proportional 
to the abundance, $\tau_{T}$ is typically an order of magnitude or more larger 
than the reaction network timestep (see, \eg, \cite{WeZW78,BeTH89}).  It is therefore
sufficient to calculate the  energy gain at the end of a timestep via
equation~\ref{eq:edot}, modified  as discussed above, and approximate
$T(t+\delta t) \approx T(t)$ or to  extrapolate based on \edot(t).  Since other
locally coupled quantities have  characteristic timescales much longer than
$\Delta t$, they too can be  decoupled in a similar fashion.  For the remainder
of this review, we will  consider only the equations governing changes in
isotopic abundances,  remembering that additional equations can easily be
constructed for those  special circumstances where they are necessary. 

Spatial coupling, particularly the modification of the composition by 
hydrodynamic movements such as diffusion, convective mixing and advection 
(in the case of Eulerian hydrodynamics methods), represents a more 
difficult challenge.  By necessity, an individual nucleosynthesis 
calculation examines the abundance changes in a locality of uniform 
composition.  The difficulties associated with strong spatial coupling of 
the composition occur because this nucleosynthetic calculation is spread 
over an entire hydrodynamic zone.  Convection can result in strong 
abundance gradients across a single hydrodynamic zone, which with the 
assumption of compositional uniformity, can result in very different 
outcomes as a function of the fineness of the hydrodynamic grid.  Thermonuclear supernovae, where the thickness of the flame front can be billionths of the radius of the white dwarf \cite{NiHi95,Khok95}, present an extreme example of microscopic nuclear inhomogeneity.  Eulerian 
advection of compositional boundaries can also have extremely unphysical 
consequences.  Fryxell \etal\ \cite{FrMA89} demonstrated how this artificial 
mixing can produce an unphysical detonation in a shock tube calculation by 
mixing cold unburnt fuel into the hot burnt region.  A related problem is 
the conservation of species.  Hydrodynamic schemes must carefully conserve 
the abundances (or partial densities) of all species \cite{FrMA89,Larr91,PlMu99}, lest they provide unphysical abundances to the nucleosynthesis calculations, which must assume conservation, and thereby produce unphysical results.  Because of these problems, hydrodynamic methods with excellent capture of shocks and contact or compositional discontinuities are best suited to nucleosynthesis calculations.

The relative size of the burning timescales, when compared to the relevant 
diffusion, sound crossing or convective timescale, dictates how these 
problems must be addressed.  If all of the burning timescales are much 
shorter than the timescale on which the hydrodynamics changes the 
composition, then the assumption of uniform composition is satisfied and 
the nucleosynthesis of each hydrodynamic zone can be treated independently.  
If all of the burning timescales are much larger than, for example, the 
convective timescale, then the composition of the entire convective zone 
can be treated as uniform and slowly evolving.  The greatest complexity 
occurs when the timescales on which the hydrodynamics and nucleosynthesis 
change the composition are similar.  Oxygen shell burning represents an 
excellent example of this as the sound travel, convective turnover and 
nuclear burning timescales are all of the same order as the evolutionary 
time.  The results of 2D simulations \cite{BaAr98} demonstrate convective 
overshooting, highly non-uniform burning and a velocity structure dominated 
by convective plumes.  In section \ref{sect:sparse} we will briefly discuss
the numerical challenges posed by fully coupling nucleosynthesis and nuclear mixing.  Silicon burning represents a different challenge \cite{Arne96}, as the timescales for the transformation of silicon to iron are much slower than the convective turnover time, but the burning timescales for the free neutrons, protons and $\alp$-particles which maintain QSE are much faster, providing a strong motivation for the 
hybrid reaction networks we will discuss in \S \ref{sect:hyb}.

\section{Solving the Nuclear Reaction Network} \label{sect:netsolve}

In principle, the initial value problem presented by the nuclear reaction network 
can be solved by any of a large number of methods discussed in the 
literature.  However the physical nature of the problem, reflected in the 
$\lambda$'s and \sigv 's, greatly restricts the optimal choice.  The large 
number of reactions display a wide range of reaction timescales, $\tau$ 
(see Eq.~\ref{eq:tau}).  Numerical systems whose solutions depend on a wide range of timescales are termed \emph{stiff}.  Gear \cite{Gear71} demonstrated that 
even a single equation can be stiff if it has both rapidly and slowly varying 
components.  Practically, stiffness occurs when the limitation of the 
timestep size is due to numerical stability rather than accuracy.  A more 
rigorous definition \cite{Lamb80} is that a system of equations $\vec{\dot Y}
(\vec Y)$ is stiff if the eigenvalues $\lambda_{j}$ of the Jacobian 
$\partial \vec{\dot Y} / \partial \vec Y$ obey the criterion that for negative $\Re(\lambda_{j})$ (the real part of the eigenvalues $\lambda_j$)
\begin{equation}
    {\mathcal S}  =  {{max |\Re(\lambda_{j})|} \over {min |\Re(\lambda_{j})|}} 
    \gg 1 
    \label{eq:lambdastiff}
\end{equation}
for $j=1,\cdots,N$.  As we will explain in this section, $\mathcal S > 10^{15}$ is not uncommon in astrophysics.

Astrophysical calculations of nucleosynthesis belong to the general field of
reactive flows, and therefore share some characteristics with related
terrestrial fields.  In particular, chemical kinetics, the study of the
evolution of chemical abundances, is an important part of atmospheric and
combustion physics and produces sets of equations much like Eq.~\ref{eq:ndot}
(see \cite{OrBo01} for a through introduction).  These chemical kinetics systems
are known for their stiffness and a great deal of effort has been expended on
developing methods to solve these equations.  Many of the considerations for
the choice of solution method for chemical kinetics also 
apply to thermonuclear kinetics.  In both cases, temporal integration of
the reaction rate equations is broken up into short intervals because of the
need to update the hydrodynamics variables.  This favors one step, self
starting algorithms.  Because abundances must be tracked for a large number of
computational cells (hundreds to thousands for one dimensional models,
millions to billions for the coming generation of three dimensional models), memory storage concerns favor low order methods since they don't require  the storage of as much data from prior steps.  In any event, both the errors in fluid dynamics and in the reaction rates are typically a few percent or more, so the greater precision of these higher order methods often does not result in greater accuracy.  Note that these statements do not in general apply to calculations of Big Bang nucleosynthesis (at least those that assume homogeniety). As a result quite different methods are employed in these calculations \cite{FiOl04}

Because of the wide range in timescales between strong, electromagnetic and 
weak reactions, nuclear reaction networks are extraordinarily stiff.  PP chain
nucleosynthesis, responsible for the energy output of the Sun, offers an
excellent example of the difficulties.  The first reaction of the PP1 chain is
$\mnuc{H}{1}(p,e^{+} \nu)  \mnuc{H}{2}$, the fusion of two protons to form
deuterium.  This is a weak  reaction, requiring the conversion of a proton into
a neutron, and  releasing a positron and a neutrino.  As a result, the reaction
timescale  $\tau_{p}(\mnuc{H}{1})$ is very long, billions of years for
conditions like  those in the solar interior.  The second reaction of the PP1
chain is the  capture of a proton on the newly formed deuteron,
$\mnuc{H}{2}(p,\gamma)  \mnuc{He}{3}$.  For conditions like those in the solar
interior, the  characteristic timescale, $\tau_{p}(\mnuc{H}{2})$ is a few
seconds.  Thus  the timescales for two of the most important reactions for
hydrogen burning  in stars like our Sun differ by more than 17 orders of
magnitude (see \cite{Clay83,HaPR04} for a more complete discussion of the PP chain). 
This  disparity results not from a lack of \mnuc{H}{1} + \mnuc{H}{1} collisions (which occur at a rate  $Y(\mnuc{H}{1})/ Y(\mnuc{H}{2}) \sim 10^{17}$ times more often than \mnuc{H}{1} + \mnuc{H}{2} collisions), but from the rarity of the transformation of a proton to a neutron.  While the presence of weak  reactions
among the dominant energy producing reactions is unique to hydrogen burning,
most nucleosynthesis calculations are similarly stiff, in part because of the
need to include weak interactions but also the potential for neutron capture
reactions, which occur very rapidly even at low temperature, following any
release of free neutrons. The nature of the nuclear reaction network  equations has thus far limited the astrophysical usefulness of the most sophisticated methods to solve stiff equations developed for chemical kinetics.  However work to harness this resource continues.  For example, Timmes \cite{Timm99} has examined higher order Kaps-Rentrop and Bader-Deuflhard semi-implicit time integration algorithms while Mott \etal\ \cite{MoOV00} have studied asymptotic and quasi-steady state methods, which do not require a matrix solution.

For a set of nuclear abundances $\vec Y$, one can calculate the time 
derivatives of the abundances, $\dot {\vec Y}$ using Eq.~\ref{eq:ydot}.   The
desired solution is the abundance at a future time, $\vec Y(t+\Delta  t)$,
where $\Delta t$ is the network timestep.  For simplicity, most past and present nucleosynthesis calculations use the simple finite difference prescription
\begin{equation}
	{{\vec Y(t+\Delta t)- \vec Y(t)} \over {\Delta t}} = (1-\Theta) \dot 
	{\vec Y}(t+\Delta t) + \Theta \dot {\vec Y}(t).
    \label{eq:deriv}
\end{equation}
This choice is also supported by the advantages of coupling low order, single step methods with hydrodynamics. With $\Theta=1$, Eq.~\ref{eq:deriv} becomes the explicit Euler method 
while for $\Theta=0$ it is the implicit backward Euler method, both of 
which are first order accurate.  For $\Theta=1/2$, Eq.~\ref{eq:deriv} is 
the semi-implicit trapezoidal method, which is second order accurate.  For 
the stiff set of non-linear differential equations which form most nuclear reaction 
networks, a fully implicit treatment is generally most successful 
\cite{ArTr69}, though the trapezoidal method has been used in Big Bang 
nucleosynthesis calculations \cite{Wago73}, where coupling to hydrodynamics 
is less important.  Solving the fully implicit version of Eq.~\ref{eq:deriv} 
is equivalent to finding the zeros of the set of equations
\begin{equation}
    \vec \calZ(t+\Delta t)\equiv {{\vec Y(t+\Delta t)- \vec Y(t)} \over 
    {\Delta t}} - \dot {\vec Y}(t+\Delta t) =0 \ .
	\label{eq:zer}
\end{equation}
This is done using the Newton-Raphson method (see, \eg, \cite{NumRec}), 
which is based on the Taylor series expansion of $\vec \calZ(t+\Delta t)$, 
with the trial change in abundances given by
\begin{equation}
	\Delta \vec Y = \left( \partial \vec \calZ (t+\Delta t) \over 
	\partial \vec Y (t+\Delta t) \right)^{-1} \vec \calZ \ ,
	\label{eq:dely}
\end{equation}
where $\partial \vec \calZ / \partial \vec Y $ is the Jacobian of $\vec 
\calZ$.  

Historically \cite{ArTr69} and in some modern applications (see, \eg, \cite{Khok96}), each timestep consists of only a single application of the procedure outlined in Eqs.~\ref{eq:zer} \& \ref{eq:dely}.  This semi-implicit backward Euler method has the advantage of a relatively small and predictable number of matrix solutions, but there are only heuristic checks that the chosen timestep results in a stable or accurate solution.  In fully implicit backward Euler schemes, iteration using the procedure of Eqs.~\ref{eq:zer} \& \ref{eq:dely} continues until both $\Delta \vec Y$ and $\vec \calZ$ are below some tolerance, providing a measure of the stability and accuracy.  If this convergence does not occur within a reasonable number of iterations, the timestep is subdivided into smaller intervals until a converged solution can be achieved, allowing the fully implicit backward Euler integration to respond to instability or inaccuracy in a way that is impossible with the semi-implicit backward Euler approach.  Higher order methods allow better estimates of the truncation error by comparing the solutions of different order schemes and sub-dividing the timestep if these errors are too large.

A potential numerical problem with the solution of Eq.~\ref{eq:zer} is the 
singularity of the Jacobian matrix, $\partial \vec \calZ (t+\Delta t)/ \partial 
\vec Y (t+ \Delta t)$.
From Eq.~\ref{eq:zer}, the individual matrix elements of the Jacobian have 
the form
\begin{equation}
	 {\partial \calZ_{i} \over \partial Y_{j}} =  
	{\delta_{ij} \over \Delta t} - {\partial \dot Y_{i} \over 
	\partial Y_{j}}\nonumber 
	= {\delta_{ij} \over \Delta t} - \sum {1 \over \tau_{j}(i)} \ ,
	\label{eq:jacobian}
\end{equation}
where $\delta_{ij}$ is the Kronecker delta, and $\tau_{j}(i)$ is the 
destruction timescale of nucleus $i$ with respect to nucleus $j$ for a 
given reaction, as defined in Eq.~\ref{eq:tau}.  The sum accounts for the 
fact that there may be more than one reaction by which nucleus $j$ is 
involved in the creation or destruction of nucleus $i$.  Along the diagonal 
of the Jacobian, there are two competing terms, $1/\Delta t$ and $\sum 
1/\tau_{i}(i)$.  This sum is over all reactions which destroy nucleus $i$, 
and is dominated by the fastest reactions.  As a result, $\sum 1/\tau_{i}(i)$ 
can be orders of magnitude larger than the reciprocal of the desired timestep, 
$1/\Delta t$.  This is especially a problem near equilibrium, 
where both destruction and the balancing production timescales are very 
short in comparison to the preferred timestep size, resulting in 
differences close to the numerical accuracy (i.e. 14 or more orders of 
magnitude).  In such cases, the term $1/\Delta t$ is numerically neglected, 
leading to numerically singular matrices.  One approach to avoiding this 
problem is to artificially scale these short, equilibrium timescales by a 
factor which brings their timescale closer to $\Delta t$, but leaves them 
small enough to ensure equilibrium.  While this approach has been used 
successfully, the ad hoc nature of this artificial scaling renders these 
methods fragile.  A more promising approach is to make directly use of 
equilibrium expressions for abundances, which, as we will discuss in \S
\ref{sect:hyb}, also ensures the economical use of computer resources.

\subsection{Taking Advantage of Matrix Sparseness}\label{sect:sparse}

For larger nuclear reaction networks, the Newton-Raphson method requires solution of a moderately large ($N=100-3000$) matrix equation for each zone.  Since general solution of a dense matrix scales as $O(N^{3})$, this can make these large networks 
progressively much more expensive.  While in principal, every species reacts 
with each of the hundreds of others, resulting in a dense Jacobian matrix, 
in practice it is possible to neglect most of these reactions.  Because of 
the $Z_{i}Z_{j}$ dependence of the repulsive Coulomb term in the nuclear 
potential, captures of free neutrons and isotopes of H and He on heavy 
nuclei occur much faster than fusions of heavier nuclei.  Furthermore, with 
the exception of the Big Bang nucleosynthesis and PP-chains, reactions 
involving secondary isotopes of H (deuterium and tritium) and He are 
neglectable.  Likewise, photodisintegrations tend to eject free nucleons or 
\alp-particles.  Thus, with a few important exceptions, for each nucleus we 
need only consider twelve reactions linking it to its nuclear neighbors by the 
capture of an $n,p,\alpha$ or $\gamma$ and release a different one of these 
four.  The exceptions to this rule are the few heavy ion reactions important 
for burning stages like carbon and oxygen burning where the dearth of light 
nuclei cause the heavy ion collisions to dominate.

\begin{figure}
    \begin{center}
 	    \includegraphics[width=.6\textwidth]{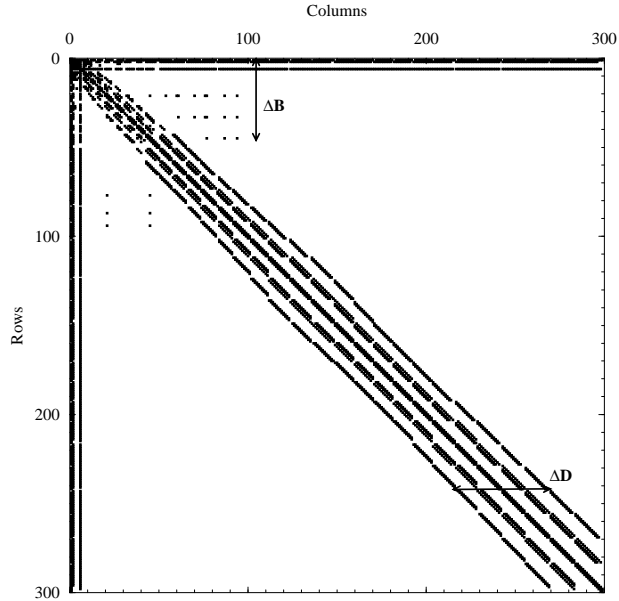}
    \end{center}
	\caption{Graphic demonstration of the sparseness of the Jacobian
	matrix.  The filled squares represent the non-zero elements.}
    \label{fig:sparse} 
\end{figure}

Fig.~\ref{fig:sparse} demonstrates the sparseness of the resulting Jacobian 
matrix, for a 300 nuclei network chosen to handle all the energy generating stages in the life of a massive star.  The nuclei are indexed in order of increasing Z and then A.  Of the 90,000 matrix elements, fewer than 5,000 are non-zero.  In terms of
the standard forms for sparse matrices, this Jacobian is best described
as doubly bordered, band diagonal.  With a border width, $\Delta B$, of
45 necessary to include the heavy ion reactions among \mnuc{C}{12}, 
\mnuc{O}{16} and \mnuc{Ne}{20} along with the free neutrons, protons and
\alp -particles and a band diagonal width, $\Delta D$, of
54, even this sparse form includes almost 50,000 elements.  With solution
of the matrix equation consuming 90+\% of the computational time, there
is clearly a need for custom tailored solvers which take better advantage 
of the sparseness of the Jacobian \cite{PrAA87}. To date best results for small (N$<$100) matrices are obtained with machine optimized dense solvers (\eg\ LAPACK) or matrix specific solvers generated by symbolic processing \cite{Muel98,Muel86}.
For large matrices, generalized sparse solvers, both custom built and from
software libraries, are used (see, \cite{Timm99}).  

Thus far in this section we have considered thermonuclear kinetics in the limit where the hydrodynamics is treated separately by operator-splitting.   However, as we discussed in \S\ref{sect:hydro}, this is approach is not always viable, so we conclude this section with some remarks on the additional challenges encountered in recent efforts to solve coupled nucleosynthesis and hydrodynamical mixing.  Because these calculations must also be finite differenced with an implicit scheme, they too require matrix solutions and all zones must be solved concurrently.  The resulting matrices are quite large: without consideration of matrix sparseness, one would need to store $N^2 M^2$ numbers, where $N$ is the number of species in each zone and $M$ is the number of zones.  Fortunately, however, the resulting matrices are quite sparse since they are block diagonal (each block is similar in form to that shown in Fig.~\ref{fig:sparse}) with bands linking adjacent spatial zones.  Such large sparse matrices are typically solved by iterative techniques, such as biconjugate gradient
which requires only matrix multiplies and, thus, only storage of the non-zero
matrix elements.  To date, computation of $N=1000$, $M=1000$ problems (that
is, $10^6 \times 10^6$ matrices) can be fairly routinely handled on a single processor.  This approach to solving coupled nucleosynthesis and nuclear mixing is necessary in cases where operator splitting is suspect, and we expect to see it finding greater use in future calculations.

\subsection{Physically Motivated Network Specialization}

Often from a physical understanding one can specialize the general solution
method and thereby greatly reduce the computational cost.  As an example,
we briefly discuss the r-process approximation described in 
\cite{CoTT91,FRRK99}.  For nuclei with $A >100$, 
charged particle captures (proton and \alp) as well as their reverse
photodisintegrations virtually cease when $T<3\gk$.  This leaves only neutron 
captures and their reverse photodisintegration reactions, as well as 
$\beta$-decays, which can also lead to the emission of delayed neutrons.  In this case, Eq.~\ref{eq:ydot} greatly simplifies, leaving 
\begin{eqnarray} 
\dot{Y}(\mnuc{Z}{A})  & = & n_n \langle \sigma v
\rangle^{n,\gamma}_{Z,A-1} Y(\mnuc{Z}{A-1})  + \lambda^{\gamma}_{Z,A+1}
Y(\mnuc{Z}{A+1})  + \sum_{j=0}^J \lambda_{Z-1,A+j}^{\beta j n}
Y(\mnuc{Z-1}{A+j}) \nonumber \\ & - & \left(n_n \langle \sigma v
\rangle^{n,\gamma}_{Z,A}+\lambda^{\gamma}_{Z,A}  + \sum_{j=0}^J
\lambda_{Z,A}^{\beta j n} \right) Y(\mnuc{Z}{A}) \ , \label{eq:rproc}
\end{eqnarray} 
where $\langle \sigma v \rangle^{n,\gamma}_{Z,A}$ and $\lambda^{\gamma}_{Z,A}$ 
are the velocity integrated neutron capture cross section and the
photodisintegration rate for the nucleus \mnuc{Z}{A}, while
$\lambda_{(Z,A)}^{\beta j n}$ is the decay constant for the $\beta^-$ decay of
\mnuc{Z}{A}, with $j$ delayed neutrons (up to a maximum of J).  The assumption is made that the neutron abundance $(Y_n=n_n/\rho N_A)$ varies slowly enough that it may be evolved explicitly.  One
can see that in Eq.~\ref{eq:rproc}, with $n_n$ thereby fixed, the time
derivatives of each species have a linear dependence on only the abundances of
their neighbors in the same isotopic chain (nuclei with the same $Z$), or that
with one less proton $(Z-1)$. One can then divide the network into separate
pieces for each isotopic chain, and solve them sequentially, beginning with the
lowest Z.  The ``boundary'' terms for this lowest $Z$ chain can be supplied by
a previously  run or concurrently running full network calculation which need
extend only  to this $Z$.  This reduces the solution of a matrix with more
than a thousand rows to the  solution of roughly 30 smaller matrices. 
Furthermore each of these smaller  matrices is also tridiagonal increasing
speed further.  Freiburghaus \etal\ \cite{FRRK99} tested  the assumption of slow variation in the neutron abundance and have demonstrated  the usefulness of this method in r-process simulations, achieving a large decrease in computational cost. 
A similar treatment has been successfully applied to explosive hydrogen burning
based on the assumption of slowly varying proton and alpha abundances
\cite{RFRT97}.  As we will discuss in  \S \ref{sect:hyb}, for other burning 
stages there also exist physically motivated simplifications to the general network 
solution method.

\section{Equilibria in Nuclear Astrophysics} \label{sect:nse}

An isolated thermodynamic system tends to evolve towards an equilibrium state,
and whether the system reaches that equilibrium depends on the competition
between the timescale for equilibration and the other relevant timescales.
For example, the reaction-rate expression found in Eq. (\ref{eq:sigv})
assumed a local thermal equilibrium in the plasma such that the velocities of
the interacting particles are well described by a Maxwell-Boltzmann
distribution, which is determined by a single parameter, the local temperature.
This is valid because the timescale to achieve this equilibrium is much shorter
than the other interaction timescales in the plasma.  Similarly, as
discussed in \S\ref{sect:reacrate}, it is usually assumed in nuclear astrophysics that the internal states of a nucleus are also in equilibrium so that the probability of finding a nucleus in particular excited state is proportional to the Boltzmann factor for that state, which, as before, is determined by the temperature and the (presumably known) excitation energy of the state.  Again this is generally a valid assumption
because the transitions within the nucleus are typically much faster than other
interactions, although interesting challenges arise when this assumption is not valid (e.g., \cite{WaFo80,GuMe01}).

At conditions of high temperature and density, the thermonuclear reaction rates
themselves may be sufficiently rapid to achieve equilibrium within the timescale
set by the hydrodynamics of the astrophysical setting.  This permits considerable simplification of the calculation of the nuclear abundances.  In most such cases, the fast strong and electromagnetic 
reactions reach equilibrium while those involving the weak nuclear force do 
not.  Since the weak reactions are not equilibrated, the resulting {\em 
Nuclear Statistical Equilibrium} (NSE) requires monitoring of weak 
reaction activity.  Even with this stricture, NSE offers many advantages, 
since hundreds of abundances are uniquely defined by the thermodynamic 
conditions and a single measure of the weak interaction history or the 
degree of neutronization.  Computationally, this reduction in the number of 
independent variables greatly reduces the cost of nuclear abundance evolution.  
Because there are fewer variables to follow within a hydrodynamic model, the 
memory footprint of the nuclear abundances is also reduced, an issue of 
importance in modern multi-dimensional models.  Finally, the 
equilibrium abundance calculations depend on binding energies and partition 
functions, quantities which are better known than many reaction rates.  
This is particularly true for unstable nuclei and for conditions where the 
mass density approaches that of the nucleus itself, resulting in exotic 
nuclear structures.

The expression for NSE is commonly derived using either chemical potentials 
or detailed balance (see, \eg, \cite{Clay83}).  
For a nucleus $^AZ$, composed of $Z$ protons and $N=(A-Z)$ neutrons, in 
equilibrium with these free nucleons, the chemical potential of $^AZ$ can 
be expressed in terms of the chemical potentials of the free nucleons
\begin{equation}
    \mu_{Z,A} = Z \mu_p + N \mu_n \ .
    \label{eq:muAZ}
\end{equation}
Substituting the expression for the Boltzmann chemical potential (including rest mass)
into Eq.~\ref{eq:muAZ} allows 
derivation of an expression for the abundance of every nuclear species in 
terms of the abundances of the free protons ($Y_p$) and neutrons ($Y_n$),
\begin{equation}
    Y(^AZ) = {G(^AZ) \over 2^A} {\left(\rho N_A \over \theta \right)}^{A-1} 
    A^{3 \over 2} \exp {\left( B(^AZ) \over {k_B T} \right)} {Y_n}^N {Y_p}^Z , 
    \label{eq:nse}
\end{equation}
where $G(^AZ)$ and $B(^AZ)$ are the partition function and binding energy 
of the nucleus $^AZ$, $N_A$ is Avagadro's number, $k_B$ is Boltzmann's 
constant, $\rho$ and $T$ are the density and temperature of the plasma, and 
$\theta = (m_u k_B T / 2 \pi \hbar^2 )^{3/2}$.

\begin{figure}
 	\includegraphics[angle=90,width=\textwidth]{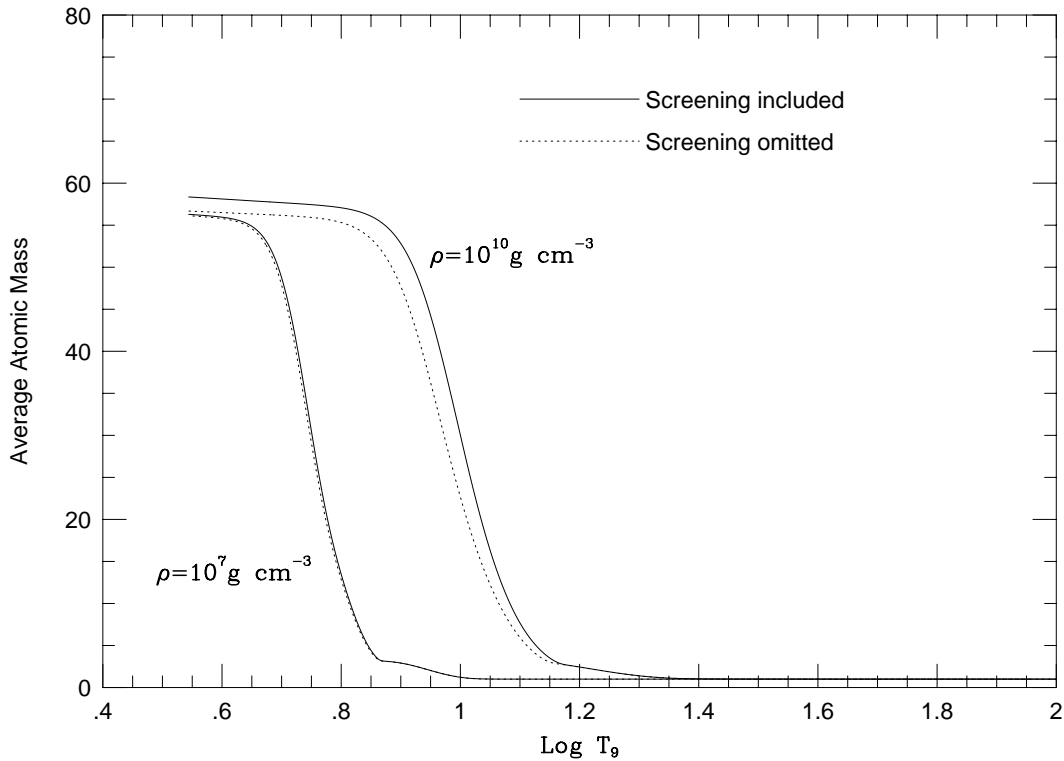}
	\caption{The average atomic mass for material in NSE as a function of 
	Temperature.  The solid lines include screening corrections to the 
	nuclear binding energies, while the dotted lines ignore this effect.}  
    \label{fig:AbarNSE} 
\end{figure}

Abundances of all nuclear species can therefore be expressed as functions of 
two quantities.  Nucleon number conservation $(\sum A Y = 1)$ provides one constraint.  The second constraint is the amount of weak reaction activity, often expressed in terms of the total proton abundance, $\sum Z Y$, which charge 
conservation requires equal the electron abundance, \ye.  Thus the nuclear 
abundances are uniquely determined for a given $(T,\rho,\ye)$.  Alternately, 
the weak interaction history is sometimes expressed in terms of the 
neutron excess $\eta=\sum(N-Z) Y$.  Figure~\ref{fig:AbarNSE} displays the 
temperature and density dependence of $\bar A = \sum A Y / \sum Y = 1/ \sum 
Y$, the average nuclear mass of the NSE distribution.  At high 
temperatures, free nucleons are favored, hence $\bar A \sim 1$.  For 
intermediate temperatures the compromise of retaining large numbers of 
particles while increasing binding energy favors \mnuc{He}{4}, which has 
$80\%$ of the binding energy of the iron peak nuclei.  At low temperatures, 
Eq.~\ref{eq:nse} strongly favors the most bound nuclei, the iron peak 
nuclei, so $\bar A \rightarrow 60$ as the temperature drops.  Density can 
be seen to scale the placement of these divisions between high, 
intermediate and low temperature and increase the average mass at low temperature.  Variations in \ye\ do not strongly 
affect Figure~\ref{fig:AbarNSE}.  At high temperatures, it simply effects 
the ratio of $Y_{p}/Y_{n} \approx \ye/1-\ye$.  At low temperatures, 
variation in \ye\ changes which Fe-peak isotopes dominate.  For example, though 
\mnuc{Ni}{56} is less tightly bound than \mnuc{Fe}{54}, it is more tightly 
bound than \mnuc{Fe}{54} + 2 \mnuc{H}{1}, which would be required by charge 
conservation if $\ye\sim .5$.  Thus $Y(\mnuc{Ni}{56}) > Y(\mnuc{Fe}{54})$ 
for low $T$ with $\ye\sim .50$, but $Y(\mnuc{Fe}{54}) > Y(\mnuc{Ni}{56})$ 
for smaller \ye.  In general, the most abundant nuclei at low temperatures 
are the most bound nuclei for which $Z/A \sim \ye$.

As with any equilibrium distribution, there are limitations on the 
applicability of NSE. For NSE to provide a good estimate of the nuclear abundances the temperature must be sufficient for the endoergic reaction of each reaction pair to occur.  Since for all particle-stable nuclei between the proton and neutron drip lines (with the exception of nuclei unstable against alpha decay), the photodisintegrations 
are endoergic, with typical Q-values among ($\beta$) stable nuclei of 8-12 
\mev, by Eq.~\ref{eq:detbal} this requirement reduces to $T>3\gk$.  While 
this requirement is necessary, it is not sufficient.  In the case of 
hydrostatic silicon burning, even when this condition is met, appreciable 
time is required to convert Si to Fe-peak elements.  In the case of 
explosive silicon burning, the adiabatic cooling on timescales of seconds 
can cause conditions to change more rapidly than NSE can follow, breaking 
down NSE first between \mnuc{He}{4} and \mnuc{C}{12}, at $T\sim 6\gk$ 
\cite{MeKC98} and later between the species near silicon and the Fe-peak 
nuclei, at $T\sim 4\gk$ \cite{WoAC73,HiTh99a}.  Thus it is clear that in the face 
of sufficiently rapid thermodynamic variations, NSE provides a problematic 
estimate of abundances.  It is important to note that, in spite of the breakdown of the global NSE, many nuclei under these conditions do obey local equilibrium relations with their neighbors.

\section{Merging Equilibria with Reaction Networks} \label{sect:hyb}

In spite of the limitations on the applicability of NSE, the reduced 
computational cost provides a strong motivation to maximize the use of 
equilibria.  The use of equilibrium expressions for single abundances is, 
in fact, common in nuclear reaction networks, typically to track the 
abundances of short-lived unstable intermediates in ``three-particle'' 
processes.  The most common example of this is the triple \alp\ process, 
\begin{equation}
    \begin{array}{rcllll}
         \mnuc{He}{4} + \mnuc{He}{4} & \rightleftharpoons & \mnuc{Be}{8} & & & Q = -.09 \mev \\
         \mnuc{He}{4} + \mnuc{Be}{8} & \rightleftharpoons & \mnuc{C^*}{12} & \rightarrow & \mnuc{C}{12} + \gamma \qquad & Q = 7.37 \mev \ ,
    \end{array}
\end{equation} 
by which Helium burning occurs.  With $\tau(\mnuc{Be}{8}) \sim 10^{-16} \sec$, 
only rarely does a \mnuc{Be}{8} survive long enough for a second \alp\ to 
capture.  As a result of the near balance of the first reaction pair, the 
abundance of \mnuc{Be}{8} can be expressed in terms of the \alp-particle 
abundance,
\begin{equation}
Y(\mnuc{Be}{8})= {\rho N_A \over \theta} \left(1 \over 2\right) ^{3/2}
\exp \left(M(\mnuc{Be}{8})-2M_{\alp}\over {k_B T}\right) Y_{\alp}^2 .
\end{equation}

Likewise for temperatures in excess of .1 \gk, the most likely result following
the second \alp\ capture to form an excited state of \mnuc{C}{12} is a decay
back to \mnuc{Be}{8} ($\Gamma_{\alp}(\mnuc{C^*}{12})/\Gamma_{\gamma}
(\mnuc{C^*}{12})>10^3$), thus the abundance of \mnuc{C^*}{12} is well
characterized by
\begin{equation}
Y(\mnuc{C^*}{12})= \left(\rho N_A \over \theta \right)^2 \left(3 \over 
16\right)^{3/2} \exp \left(M(\mnuc{C^*}{12})-3M_{\alp}\over {k_B T}\right) 
Y_{\alp}^3 .
\end{equation}

When this is the case, the effective triple \alp\ reaction rate is
simply that of the decay of \mnuc{C}{12} from the excited state to the
ground state,
\begin{equation}
r_{3\alp} = \rho N_A Y(\mnuc{C*}{12}) \Gamma_{\gamma}(\mnuc{C^*}{12}) / \hbar \ .
\end{equation}

This use of local equilibrium within a rate equation shares many characteristics with the more elaborate schemes we will discuss next. The number of species tracked by the network is reduced since $Y(\mnuc{Be}{8})$ need not be directly evolved.  Problematically small timescales like $\tau(\mnuc{Be}{8})$ are removed, replaced by larger time scales ($\tau_{3\alp} \sim 10^5-10^7$ years during core helium burning).  The non-linearity of network time derivatives is increased ($\dot Y(\mnuc{C}{12})
\propto Y_{\alp}^3$) under this scheme.  This approximation also breaks down 
at low $T$ (for details see \cite{NoTM85}). 

\section{The QSE-reduced Network} \label{sect:qse}

As we noted in the previous section, while global NSE may not always apply for temperatures in the range 3-6 \gk, many nuclei are in local equilibrium with their neighbors.  Beginning with Bodansky \etal\ \cite{BoCF68}, a number of attempts
have  been made to take advantage of these partial equilibria (termed quasi-equilibria or QSE) to reduce the number of independent variables evolved via rate equations and thereby  reduce the computational cost of modeling these burning
stages.  To evolve the abundances of every member of a QSE group, it is sufficient to evolve the abundance of any single group member along with the abundances of the free nucleons.  One can thereby reduce the number of abundances that are
evolved, while still calculating, from QSE relations, the abundances of all members of a QSE group and the resultant reaction rates, including the electron and neutrino capture reactions responsible for changes in the neutronization. The result is a more computationally efficient method that retains the accuracy of the full network and yields abundances for all nuclei found in the full network.  Such methods have been applied to the \alp-network and produced networks that are twice as fast as the minimal \alp-chain network, without significantly affecting the nuclear evolution \cite{HKWT98,TiHW00}.  Reductions of an order of magnitude in computational cost have been achieved \cite{FrHT99,HiFT04} for QSE-reduced networks of the size necessary to capture the essential features of supernova nucleosynthesis and we believe greater savings are possible with further refinement.  In addition to silicon burning, there  are a number of astrophysically important situations where $T > Q/30k_{B}$  for at least some of the relevant reactions and so large equilibrium  groups exist, but NSE is not globally valid.  This includes the r-process \cite{CoTT91} and the rp-process in X-ray bursts (and possibly novae) \cite{RFRT97}, where neutron or proton separation energies ($Q_n$ or $Q_p$) of 2~\mev and less are  often encountered.  

As a example, we will discuss the QSE-reduced \alp-network.  Its mission is to evolve the abundances of the full 14 elements of a conventional \alp-network (which we'll call \yf), and calculate the resulting energy generation, in a more efficient way.  Under conditions where QSE applies, the existence of the silicon and iron peak 
QSE groups (which are separated by the nuclear shell closures Z=N=20 and 
the resulting small Q-values and reaction rates) allows calculation of 
these 14 abundances from 7.  For the members of the silicon group 
(\mnuc{Si}{28}, \mnuc{S}{32}, \mnuc{Ar}{36}, \mnuc{Ca}{40}, \mnuc{Ti}{44}) 
and the iron peak group (\mnuc{Cr}{48}, \mnuc{Fe}{52}, \mnuc{Ni}{56}, 
\mnuc{Zn}{60}) the individual abundances can be calculated by expressions 
similar to Eq.~\ref{eq:nse},
\begin{eqnarray}
    Y_{QSE,\mathrm{Si}}(^AZ) & = & {{C(^AZ)}\over{C(\mnuc{Si}{28})}} 
    Y(\mnuc{Si}{28}) Y_{\alpha}^{{A-28} \over 4} \nonumber  \\
    Y_{QSE,\mathrm{Ni}}(^AZ) & =  & {{C(^AZ)}\over{C(\mnuc{Ni}{56})}} 
    Y(\mnuc{Ni}{56}) Y_{\alpha}^{{A-56} \over 4},
    \label{eq:yq}
\end{eqnarray}
where ${C(^AZ)}$ is defined in Eq.~\ref{eq:nse} and $(A-28)/4$ and 
$(A-56)/4$ are the number of \alp-particles needed to construct $^{A}Z$ 
from \mnuc{Si}{28} and \mnuc{Ni}{56}, respectively.  Where QSE 
applies, \yf\ is a function of the abundances of a reduced nuclear set, \calR, 
defined as \alp, \mnuc{C}{12}, \mnuc{O}{16}, \mnuc{Ne}{20}, \mnuc{Mg}{24}, 
\mnuc{Si}{28}, \mnuc{Ni}{56} and we need only evolve \yr.  It should be 
noted that \mnuc{Mg}{24} is ordinarily a member of the silicon QSE group 
\cite{Arne96,WoAC73,HiTh96}, but for easier integration of 
prior burning stages with a conventional nuclear network, we will evolve 
\mnuc{Mg}{24} independently.  The main task when applying such hybrid 
schemes is finding the boundaries of QSE groups and where individual 
nuclei have to be used instead.  Treating marginal group members as part of 
a group increases the efficiency of the calculation, but may decease the 
accuracy.

While \yr\ is a convenient set of abundances for calculating \yf, it is not 
the most efficient set to evolve, primarily because of the non-linear 
dependence on $Y_{\alpha}$.  Instead we define a set of group abundances, \yg = $[Y_{\alpha G}$, 
$Y(\mnuc{C}{12})$, $Y(\mnuc{O}{16})$, $Y(\mnuc{Ne}{20})$, $Y(\mnuc{Mg}{24})$, 
$Y_{SiG}$, $Y_{FeG}]$ where
\begin{eqnarray}
    Y_{\alp G} & = & \quad Y_{\alpha} + \sum_{i \in Si \ group} 
	{{A_{i}-28} \over 4} Y_{i} + \sum_{i \in Fe \ group} {{A_{i}-56} \over 
	4} Y_{i} \ , \nonumber \\
    Y_{Si G} & = & \sum_{i \in Si \ group} Y_{i} \ , \label{eq:yg} \\
    Y_{Fe G} & = & \sum_{i \in Fe \ group} Y_{i} \ . \nonumber 
\end{eqnarray}
Physically, $Y_{\alpha G}$ represents the sum of the abundances of free 
\alp-particles and those \alp-particles required to build the members of the 
QSE groups from \mnuc{Si}{28} or \mnuc{Ni}{56}, while $Y_{Si G}$ and $Y_{Fe
G}$  represent the total abundances of the silicon and iron peak QSE groups.
This method, which here is applied only to the chain of \alp-nuclei can also
be generalized to arbitrary networks \cite{HiFT04}.  For larger networks 
which contain nuclei with $N\neq Z$, one must be able to follow the abundances
of free neutrons and protons, particularly since weak interactions will change
the global ratio of neutrons to protons.  In place of $Y_{\alpha G}$ in
Eq.~\ref{eq:yg}, one constructs  $Y_{NG}= \sum_{i,light} N_i Y_i+ \sum_{i,Si}
(N_i-14) Y_i +  \sum_{i,Fe} (N_i-28) Y_i$ and $Y_{ZG}= \sum_{i,light} Z_i Y_i
+ \sum_{i,Si} (Z_i-14) Y_i + \sum_{i,Fe} (Z_i-28) Y_i $, if \mnuc{Si}{28} and
\mnuc{Ni}{56} are chosen as the focal nuclei for the Si and Fe groups.  

Corresponding to this reduced set of abundances \calG\ is a reduced set of 
reactions, with quasi-equilibrium allowing one to ignore the reactions among 
the members of the QSE groups. Unfortunately, the rates of these remaining 
reactions are functions of the full abundance set, \yf, and are not easily 
expressed in terms of the group abundances, \yg.  Thus, for each \yg, one must 
solve for \yr\ and, by  Eq.~\ref{eq:yq}, \yf, in order to calculate \ygdot\
which is needed to evolve \yg\ via Eq.~\ref{eq:deriv}.  Furthermore,
Eq.~\ref{eq:dely} requires the calculation of the Jacobian of $\vec \calZ$, 
which can not be calculated directly since \ygdot\ cannot be expressed in 
terms of \yg.  Instead it is sufficient to use the chain rule,
\begin{equation}
	{{\partial \ygdot} \over {\partial \yg}} = {{\partial \ygdot} \over 
	{\partial \yr}} {{\partial \yr} \over {\partial \yg}}	
	\label{eq:chain}
\end{equation}
to calculate the Jacobian.  Analytically, the first term of the chain rule 
product is easily calculated from the sums of reaction terms, while the 
second term requires implicit differentiation using Eq.~\ref{eq:yg}.  Additional details and comparisons with the full \alp-network are demonstrated by Hix \etal\ \cite{HKWT98} (See also \cite{TiHW00}).

\section{Conclusion}

Computational astrophysics is in the midst of a paradigm shift. For many years, one dimensional Lagrangian models, which take advantage of the (near) sphericity of self-gravitating fluids, have been the standard for modeling the lives, deaths and afterlives of stars.  Increasingly, we see these models being supplanted by multi-dimensional models which can more accurately include the wide variety of phenomena which break spherical symmetry (for example, rotation, convection and/or magnetic fields). Present day computational capability limits the use of such multi-dimensional models to short-lived episodes in the stellar life cycle, but the future will see increasing usage of such models.  An important limitation of the current multi-dimensional models is the small number of nuclear species which are evolved within these models, typically less than twenty.  This is in contrast to the hundreds or thousands of species tracked in the one dimensional models.  Removing this limitation is vital if we are to continue to compare models to the myriad of observations which reveal the elemental and isotopic composition of the Universe.  This will require continued improvements in the numerical methods we have discussed.

In this article, we have reviewed the numerical methods presently used to study thermonuclear kinetics in astrophysics and highlighted the directions in which we see promise for improvements in speed and accuracy.  Nuclear compositions are currently evolved within astrophysical simulations via thermonuclear rate equations or Nuclear Statistical Equilibrium.  While NSE solutions are much more economical, principally because of the much smaller number of free parameters that must be evolved, they are applicable to only a few situations.  The principle difficulty with evolution via rate equation is the computational cost, which results 
primarily from the stiffness of the system of nuclear reactions.  This 
extreme stiffness requires implicit solution and has thus far generally 
precluded the use of integration methods that do not rely on the Jacobian 
matrix.  Such non-Jacobian methods remain highly sought after as a means to 
reduce the computational cost of nucleosynthesis calculations.  For 
Jacobian based integration methods, there remain considerable economies to 
be gained by taking better advantage of the sparse nature of the Jacobian and by reuse of the inverted Jacobian.

Physically motivated approaches can also be extremely useful in reducing the 
computational cost.  One such is the use of local equilibria to reduce the 
size of the system of rate equations (and reduce problems with matrix 
singularity). Methods based on local equilibria are applicable to many situations where global equilibrium has not been achieved.  Though the use of hybrid equilibrium networks is in its infancy, it seems that in many of the situations where one would have heretofore used a large network coupled with hydrodynamics, the hybrid equilibrium networks provide sufficient accuracy and considerable reduction 
in the computational cost.  Ultimately, both physically inspired  and computationally motivated improvements will be necessary as we strive to extend the reach of multi-dimensional simulations.

\begin{ack}

As a review, this article naturally owes much to the innumerable 
investigators who have devoted at least part of their life's work to our 
understanding of nucleosynthesis in astrophysical environments.  The 
authors would like to specifically thank W.D. Arnett, G. Bazan, A.G.W.
Cameron, E. M\"uller, K. Nomoto, T. Plewa, F.-K. Thielemann and S.E. Woosley who were especially influential in the preparation of this review.  Comments from the referee, F.X. Timmes, were particularly helpful.  The work has been partly supported by the National Science Foundation under contracts AST-9877130, AST-9819877, and PHY-0244783, by the Department of Energy, through the Scientic Discovery through Advanced Computing Programs, and by funds from the Joint Institute for Heavy Ion Research at ORNL.  Oak Ridge National Laboratory is managed by UT-Battelle, LLC, for the U.S. Department of Energy under contract DE-AC05-00OR22725.
\end{ack}

%\bibliographystyle{elsart-num} 
%\bibliography{apjmnemonic,apj_add,hix,nova,rspn_process,sn,network}

\end{document}